\documentclass[11pt,a4paper]{article}

\usepackage[ansinew]{inputenc}
\usepackage{amsmath, amssymb, graphics, amsthm}
\usepackage{epsfig}
\usepackage{color, xcolor}
\usepackage{fancyhdr} 
\usepackage{manfnt}
\usepackage[T1]{fontenc}
\usepackage{hyperref}

\usepackage[normalem]{ulem}

\newcommand{\ket}[1]{\left| #1 \right\rangle}

\makeatletter
\@addtoreset{equation}{section}
\makeatother

\newcommand{\be}{\begin{equation}}
\newcommand{\ee}{\end{equation}}
\newcommand{\ba}{\begin{eqnarray}}
\newcommand{\ea}{\end{eqnarray}}

\def\pb#1{\rlap{\lower1.5ex\hbox{$\longleftarrow$}}{#1}}
\def\dpb#1{\rlap{\lower1.5ex\hbox{$\Longleftarrow$}}{#1}}
\def\spb#1{\rlap{\lower1.0ex\hbox{$\leftarrow$}}{#1}}
\def\sdpb#1{\rlap{\lower1.0ex\hbox{$\Leftarrow$}}{#1}}

\title{{\sf The algebra of observables in Gau{\ss}ian normal spacetime coordinates}}
\author{
{\sf N. Bodendorfer}$^{1}$\thanks{{\sf 
norbert.bodendorfer@fuw.edu.pl}},
{\sf P. Duch}$^{2}$\thanks{{\sf 
pawel.duch@uj.edu.pl}},
{\sf J. Lewandowski}$^{1}$\thanks{{\sf 
jerzy.lewandowski@fuw.edu.pl}},
{\sf J. \'Swie\.zewski}$^1$\thanks{{\sf 
swiezew@fuw.edu.pl}}\\
\\
{\sf$^1$  Faculty of Physics, University of Warsaw, Pasteura 5, 02-093, Warsaw, Poland}\\
{\sf$^2$ Institute of Physics, Jagiellonian University, \L ojasiewicza 11, 30-348 Krak\'ow, Poland}}
\date{{\small\sf \today}}

\begin{document} 

\maketitle

{\sf
\abstract{
We discuss the canonical structure of a spacetime version of the radial gauge, i.e. Gau{\ss}ian normal spacetime coordinates. While it was found for the spatial version of the radial gauge that a ``local'' algebra of observables can be constructed, it turns out that this is not possible for the spacetime version. The technical reason for this observation is that the new gauge condition needed to upgrade the spatial to a spacetime radial gauge does not Poisson-commute with the previous gauge conditions. It follows that the involved Dirac bracket is inherently non-local in the sense that no complete set of observables can be found which is constructed locally and at the same time has local Dirac brackets. A locally constructed observable here is defined as a finite polynomial of the canonical variables at a given physical point specified by the Gau{\ss}ian normal spacetime coordinates.
}
}

\tableofcontents

\section{Introduction}

The construction of Dirac observables in theories with gauge redundancy is a fundamentally important subject of research, since only Dirac observables allow one to extract the ``physical'', i.e. gauge invariant, predictions of the theory. An alternative, but equivalent route is to introduce gauge fixings and employ the associated Dirac brackets for the computation of the dynamics. At the Hamiltonian level, the gauge redundancy translates in the appearance of constraints, and a rigorous mathematical framework to deal with them is available \cite{DiracLecturesOnQuantum, HenneauxQuantizationOfGauge}, including explicit formulas for the construction of Dirac observables once the related gauge conditions are specified \cite{AnishettyGaugeInvarianceIn, DittrichPartialAndComplete, DaporRelationalEvolution}. 

In general relativity, the gauge redundancy is given by the invariance of the theory under spacetime diffeomorphisms. Consequently, in order to talk about a certain field at a point in an invariant way, we need to specify this point via some relational construction \cite{RovelliWhatIs}. In other words, we would choose four independent fields (or more general phase space functions) with suitably non-vanishing gradient and prescribe our point as the point at which these fields take a certain value. Clearly, such a construction defines a physical coordinate system and an observable defined as some other field at this point is diffeomorphism invariant. 
Various examples for such constructions have been given in the literature dating back to \cite{DeWittTheQuantizationOf}, e.g. using dust \cite{BrownDustAsStandard}, a perfect fluid \cite{GornickaHamiltonianTheoryOf}, scalar fields \cite{RovelliThePhysicalHamiltonian}, geodesics (Gau{\ss}ian normal coordinates) \cite{DuchObservablesForGeneral, HeemskerkConstructionOfBulk, KabatDecodingTheHologram, DonnellyDiffeomorphismInvariantObservables}, and many more. 

Given a certain construction of observables, the most important questions to answer are about the algebra of the Dirac observables and the physical Hamiltonian. On the one hand, one would like to know the algebra of observables in order to be able to quantise the theory. If the algebra turns out to be complicated, e.g. non-local\footnote{Contrary to what is sometimes claimed or suggested, complete (= separating in the reduced phase space) algebras of observables may very well be local, even if the gauges involved are purely geometric in nature \cite{BSTI} or relational with respect to a non-local object, e.g. a geodesic \cite{DuchObservablesForGeneral, DuchAddendumObservablesFor}, see also section \ref{sec:Comments}, comment \ref{com:Gauges}.}, one would expect little chance of finding a Hilbert space representation thereof. On the other hand, the physical Hamiltonian which generates the time evolution with respect to the chosen clock field should be sufficiently simple and explicitly known (which may require the inversion of some differential operator), again also to be able to quantise it. 
Another interesting question that can be asked is about the local quantum field theory limit of a theory of quantum gravity, see e.g. \cite{DonnellyDiffeomorphismInvariantObservables, GiddingsHilbertSpaceStructure} for a recent discussion. If we choose to perform a reduced phase space quantisation, different choices of Dirac observables will not be equally well suited for establishing such a limit. In particular, one would like to maintain a suitable form of locality of at least the non-gravitational sector of the theory.  

Similar questions have recently also become relevant within the context of the AdS/CFT correspondence \cite{MaldacenaTheLargeN, GubserGaugeTheoryCorrelators, WittenAntiDeSitter}. There, the reconstruction of bulk fields from boundary operators has to deal with the question of how bulk points are invariantly specified. Recently, the spacetime radial gauge was employed for this \cite{HeemskerkConstructionOfBulk, KabatDecodingTheHologram, DonnellyDiffeomorphismInvariantObservables}, see also \cite{AlmheiriBulkLocalityAnd} for a related construction. In this line of work, it is a crucial question whether the bulk matter fields commute, and contradictory claims were made \cite{KabatDecodingTheHologram, DonnellyDiffeomorphismInvariantObservables}. 

In this paper, we will reinvestigate the issue of a spacetime version of the radial gauge, building on the results previously derived for a purely spatial version of it \cite{DuchObservablesForGeneral, BLSII}. We will work completely non-perturbatively, as opposed to the perturbative discussion in \cite{DonnellyDiffeomorphismInvariantObservables}. Our results will back those of \cite{DonnellyDiffeomorphismInvariantObservables}, i.e. the bulk matter fields do not commute at spacelike separation. The underlying reason for this will be identified as a the non-commutativity of the four gauge conditions underlying the spacetime radial gauge. Our discussion should thus settle similar questions by a simple check of the commutativity properties of the chosen gauge conditions. 

This paper is organised as follows: We recall the spatial version of the radial gauge as an introduction to the subject in section \ref{sec:SpatialRadialGauge}. The main part of the paper is section \ref{sec:SpacetimeRadialGauge}, where the spacetime radial gauge is implemented and the non-locality of the algebra of observables is exposed. We make some comments in section \ref{sec:Comments} and conclude in section \ref{sec:Conclusion}.

\section{The spatial radial gauge}
\label{sec:SpatialRadialGauge}

We will first briefly review the spatial version of the radial gauge, which will serve as a foundation for what follows. While the idea of using a radial\footnote{Also referred to as axial gauge \cite{DonnellyDiffeomorphismInvariantObservables}, holographic gauge \cite{KabatDecodingTheHologram}, or Fefferman-Graham gauge \cite{MintunBulkBoundaryDuality}.} gauge is quite old, the algebra of the associated observables for the spatial version has only been computed recently \cite{DuchObservablesForGeneral}. Based on this construction of observables, a reduced phase space formulation was established in \cite{BLSII}. 

The main idea is to specify coordinates on the spatial slice via the endpoints of spatial geodesics anchored either at a central point or at spatial infinity. From a gauge fixing perspective, one would demand that in a given coordinate system where the radial direction is denoted by $r$ and the angular directions by $A$, compoundly written as $a = r,A$, the spatial Christoffel symbols  
\be
	\stackrel{(3)}{\Gamma} {}^A_{rr} = \frac{1}{2} q^{Aa} \left( 2 q_{ra,r} - q_{rr,a}\right)  \label{eq:Gamma3A} 
\ee
vanish. One can furthermore demand that also
\be
	\stackrel{(3)}{\Gamma} {}^r_{rr} = \frac{1}{2} q^{ra} \left( 2 q_{ra,r} - q_{rr,a}\right) \label{eq:Gamma3r} 
\ee
vanishes, which means that the coordinate $r$ measures (a multiple of the) proper distance along the geodesics. While one could directly impose \eqref{eq:Gamma3A} = 0 and \eqref{eq:Gamma3r} = 0 as gauge conditions, which is equivalent to $2 q_{ra,r} - q_{rr,a} = 0$, it is more convenient to impose the stronger condition 
\be
	q_{ra} = \delta_{ra},
\ee
which is easier\footnote{While it seems hard to avoid setting $q_{rA} = 0$ in practical applications, it might be advantageous not to fix $q_{rr} = 1$ classically, but to only gauge fix $q_{rr,A} = 0$ classically and solve the remaining parts of the radial diffeomorphisms at the quantum level, as done in \cite{BZI}.} to deal with in practise and fleshes out the geometric interpretation of the gauge.

$q_{ra} = \delta_{ra}$ can now be interpreted as a gauge fixing condition for the spatial diffeomorphism constraint and the Dirac brackets can be constructed, as done in \cite{BLSII}. It is important to notice that the gauge conditions $q_{ra} = \delta_{ra}$ are mutually Poisson-commuting, which leads to a Dirac bracket $\{f, g\}_{\text{DB}}$ where all corrections with respect to the Poisson bracket $\{f, g\}$ are proportional to $\{f, q_{ra}\}$ or $\{g, q_{ra}\}$ due to the general inversion formula 
\be\label{InverseDiracMatrixIntro}
	\left( \begin{array}{cc}
A& B   \\
C & 0  \end{array} \right)^{-1}  = 
\left( \begin{array}{cc}
0& C^{-1}   \\
B^{-1} & -B^{-1}AC^{-1}  \end{array} \right)
\ee
for a (Dirac) matrix of the given block form.\footnote{A more detailed discussion of the Dirac bracket and the Dirac matrix is presented in section \ref{sec:DiracBracket} and appendix \ref{DiracBracketFullTheory}. The purpose of equation \eqref{InverseDiracMatrixIntro} is to indicate, that vanishing of the bottom-right block, describing the brackets between the gauge conditions, leads to a vanishing of the top-left block in the inverse. This has profound consequences for the algebra of observables in the gauge-fixed theory as described in the rest of the current paper.} All fields Poisson-commuting with $q_{ra}$, in particular additional matter fields, thus have Dirac brackets identical to their Poisson brackets. If $P^{ra}$ is involved on the other hand, the Dirac brackets are non-trivial and given in \cite{BLSII}. However, we can solve for $P^{ra}$ via the spatial diffeomorphism constraint in terms of variables having canonical Poisson brackets, that is $q_{AB}$, $P^{AB}$, and the matter fields. We thus can find a complete set of observables which have canonical, in particular local, Poisson brackets (or equivalently a complete set of fields with canonical Dirac brackets). These observables are constructed from fields living at a point which is specified by the mentioned geometric construction, and are thus local in this sense\footnote{The Hamiltonian however contains non-localities which originate from solving the spatial diffeomorphism constraint for $P^{ra}$, see also section \ref{sec:Comments}, comment \ref{comment:QFT}.}.

The spatial radial gauge can also be combined with an additional gauge condition for the Hamiltonian constraint. In \cite{DuchObservablesForGeneral, BLSII} this was tentatively specified as deparametrisation with respect to non-rotating dust \cite{BrownDustAsStandard, HusainTimeAndA}. The dust clock trivially Poisson-commutes with $q_{ra}$ and thus the two gauge fixings are effectively independent of each other. However, in the case of the spacetime radial gauge, the new gauge fixing condition does not Poisson commute with $q_{rr}$, which leads to the problem discussed in the next section.

\section{The spacetime radial gauge}
\label{sec:SpacetimeRadialGauge}

In order to upgrade the spatial radial gauge to a spacetime gauge, we need to study the form of the spacetime Christoffel symbols and translate the conditions for a radial geodesic to remain a radial (but now spacetime) geodesic into the canonical framework. Similar discussions are found in \cite{KabatDecodingTheHologram, DonnellyDiffeomorphismInvariantObservables}.

We start with the usual ADM decomposition \cite{ArnowittTheDynamicsOf}
\be
	g_{\mu \nu} =  \left( \begin{array}{cc}
-N^2 + N^a N_a & N_a   \\
N_a & q_{ab}  \end{array} \right) 
~~~~~~~
g^{\mu \nu} =  \left( \begin{array}{cc}
-1/N^2 & N^a / N^2  \\
N^a / N^2 & q^{ab} - N^a N^b / N^2  \end{array} \right),
\ee
where the spatial indices are raised and lowered with the use of spatial metric and its inverse, and obtain after a few lines of computation
\ba
	\stackrel{(4)}{\Gamma} {}^t_{rr} &= & \frac{1}{N}K_{rr}\label{eq:Gamma4t}, \\
	\stackrel{(4)}{\Gamma} {}^r_{rr} &=& -N^r \stackrel{(4)}{\Gamma} {}^t_{rr}   + \frac{1}{2 } q^{rb} \left(  2 q_{rb,r} - q_{rr,b} \right)  \text{,} \label{eq:Gamma4r}\\
	\stackrel{(4)}{\Gamma} {}^A_{rr} &=& -N^A \stackrel{(4)}{\Gamma} {}^t_{rr}   + \frac{1}{2 } q^{Ab} \left(  2 q_{rb,r} - q_{rr,b} \right)  \text{,} \label{eq:Gamma4A}
\ea
where $K_{ab}=\frac{1}{2N}(q_{ab,t}-2\stackrel{(3)}{\nabla}_{(a}\!N_{b)})$ is the extrinsic curvature tensor. In case we take our gravitational theory to be general relativity, we have the ADM \cite{ArnowittTheDynamicsOf} Poisson structure $\{q_{ab}(x), P^{cd}(y)  \} = \delta^{(3)}(x,y) \delta_{(a}^{c} \delta_{b)}^{d}$ with $P^{ab} = \frac{1}{2 \kappa} \sqrt{q} \left(K^{ab} - q^{ab} K \right)$ and the constraints given by
\ba
	H[N] &=& \int d^3x N \left( \frac{ 2\kappa}{\sqrt{q}} \left( P^{ab} P_{ab} - \frac{1}{2} P^2 \right) - \frac{ \sqrt{q}}{2 \kappa} R^{(3)}\right),\\
	C_a[N^a] &=& -2 \int d^3x N^a \nabla_b P^{b} {}_a \, \text{.}
\ea

In order to have both \eqref{eq:Gamma4t} and \eqref{eq:Gamma4A} vanishing, it is necessary that also $F^A := q^{Ab} \left(  2 q_{rb,r} - q_{rr,b} \right)$ vanishes. Since $K_{rr}=\frac{2\kappa}{\sqrt{q}}(P_{rr}-\frac{1}{2}q_{rr}P)$, it follows that $\left\{ F^A, \stackrel{(4)}{\Gamma} {}^t_{rr} \right\}  \neq 0$, which will lead to the main conclusion of the paper. However, instead of being completely general in our presentation, we proceed as in the spatial case and not only demand that \eqref{eq:Gamma4t} and \eqref{eq:Gamma4A} vanish, but also that \eqref{eq:Gamma4r} vanishes and furthermore
\be\label{GaugeConditionMetric}
	q_{ra} = \delta_{ra}.
\ee
We stress that both \eqref{eq:Gamma4r} = 0 and $q_{ra} = \delta_{ra}$ are subsequently stronger conditions than just \eqref{eq:Gamma4t} and \eqref{eq:Gamma4A}, however they do not interfere with the main conclusion of this paper and allow for a straightforward geometric interpretation.

From $q_{ra} = \delta_{ra}$, it now follows that $2 q_{rb,r} - q_{rr,b}=0$, whereas from \eqref{eq:Gamma4t}=0 we see that 
\be\label{GaugeConditionCurvature}
	K_{rr} = 0
\ee
has to be added as our fourth gauge condition. We note for later use that this can be written more conveniently, using $q_{rA} = 0$, as
\be
	P^{rr} q_{rr}  - P^{AB} q_{AB} = 0 \label{eq:ConformalGenerator} \text{,}
\ee
which is just a generator of local scale transformations on the gravitational variables (with different weights for the radial and angular components\footnote{We note here, that it is particularly transparent in a more general setting of $D$ spatial dimensions where \eqref{eq:ConformalGenerator} takes the form
\be
	(D-2) P^{rr} q_{rr}  - P^{AB} q_{AB} = 0.
\ee
}). 

One would now like to proceed as above and compute the set of phase space functions which Poisson-commute with the gauge conditions, and use them to construct a local algebra of observables. This however fails, essentially because the gauge conditions are not Poisson-commuting with each other. To see this explicitly, let us compute the Dirac bracket.

\subsection{Spherically symmetric setting}\label{sec:DiracBracket}

The key feature of the Dirac bracket which is needed for the argument discussed here is already present in the spherically symmetric setting. Since it is much more transparent there, we will discuss the case of spherically symmetric general relativity coupled to a scalar field in this section. We will do so employing the midisuperspace approach developed in \cite{KucharGeometrodynamicsOfThe, RomanoSpherically}. 
The computation of the Dirac bracket for the full theory is presented in Appendix \ref{DiracBracketFullTheory}.

\subsubsection{Setup}

In spherically symmetric midisuperspace models, the spatial line element is given by
\begin{equation}
	ds^2 = \Lambda^2 dr^2 + R^2 d\Omega^2,
\end{equation}
where $\Lambda^2$ corresponds to the $q_{rr}$ component of the metric of the non-symmetric theory, while, due to the symmetry, the $q_{rA}$ components are automatically zero (we of course center the spherical coordinate system in the center of symmetry). The remaining variable $R$ describes the areas of spheres at a given distance $r$ from the center of symmetry, which means $4\pi R^2$ corresponds to $\int d\Omega \sqrt{\det q_{AB}}$. The canonically conjugate momenta are denoted by $P_\Lambda$ and $P_R$ and the Poisson brackets read
\begin{equation}
	\{R(r),\ P_R(\bar r)\} = \delta(r-\bar r),\qquad \{\Lambda(r),\ P_\Lambda(\bar r)\} = \delta(r-\bar r).
\end{equation}
We consider a coupling to a scalar field, so we also have the canonical pair
\begin{equation}
	\{\phi(r),\ P_\phi(\bar r)\} = \delta(r-\bar r).
\end{equation}
The Hamiltonian of the theory is a sum of a vector and Hamiltonian constraint, which are given by
\begin{subequations}\label{SSconstraints}
\begin{align}
	C[N^r] &= \int dr N^r c(r) = \int dr N^r \left(P_R R' - \Lambda P'_\Lambda + P_\phi\phi' \right),\label{SSVectorConstraint}\\
	H[N] &= \int dr N h(r) = \int dr N  \left(\frac{1}{\chi}\big(\frac{\Lambda P_\Lambda^2}{2 R^2} - \frac{P_R P_\Lambda}{R}\big) + \chi\big(\frac{R R''}{\Lambda} - \frac{R R' \Lambda'}{\Lambda^2} + \frac{R'^2}{2 \Lambda} - \frac{\Lambda}{2}\big) + h^\text{matt}\right),\label{SSHamiltonianConstraint}
\end{align}
\end{subequations}
where the shift vector has only a radial component due to the symmetry requirement, the prime denotes a radial derivative, $\chi$ is a coupling parameter\footnote{To restore the units convention $G=1=c$ used in \cite{KucharGeometrodynamicsOfThe} one should put $\chi = 1$ and to restore the units convention commonly used in quantum gravity literature, namely $8\pi G = 1=c$ one should put $\chi = 8\pi$. Within the current paper, in the context of spherical symmetry, we will keep the constant $\chi = \frac{1}{G}$, while in the context of the full theory, we introduced the constant $\kappa=8 \pi G$.} and the $h^\text{matt}$ term depends on the type of scalar field one considers as well as the cosmological constant. We will assume that $\phi$ is minimally coupled, which means $h^\text{matt}$ is a functional of $\phi$, $P_\phi$, $\Lambda$ and $R$ but does not depend on the gravitational momenta $P_\Lambda$ and $P_R$, and also not on the radial derivatives of $\Lambda$ and $R$. Since we are working in spherically symmetric context, all the fields considered are functions of at most the radial coordinate $r$ and time, but they do not depend on the angles.

The radial-radial component of the extrinsic curvature, $K_{rr}$, can be expressed in terms of the above variables in the following way\footnote{Note, that our convention for the definition of the extrinsic curvature differs from the one used in \cite{KucharGeometrodynamicsOfThe} by a sign factor.}
\begin{equation}
	K_{rr} = \frac{1}{\chi}\left(\frac{\Lambda^2 P_\Lambda}{R^2} - \frac{\Lambda P_R}{R}\right).
\end{equation}

A convenient notation we will use later is to introduce
\begin{equation}
	m_\text{ADM} := \frac{1}{2}\left(\frac{P_\Lambda^2}{\chi R} + \chi R\left(1 - \frac{{R'}^2}{\Lambda^2}\right)\right)
\end{equation}
which is the (local) expression of the ADM mass of a vacuum spacetime (see \cite{KucharGeometrodynamicsOfThe} for a discussion of its properties).

\subsubsection{Spacetime radial gauge and Dirac bracket}

Imposing the spatial radial gauge in the spherically symmetric context amounts to demanding
\begin{equation}\label{SSLambdaConstraint}
	\Lambda - 1 = 0,
\end{equation}
which has been extensively discussed in \cite{BLSII}. Requiring that the radial geodesics are also spacetime geodesics amounts to requiring additionally
\begin{equation}\label{SSKrrConstraint}
	\frac{1}{\chi}\left(\frac{\Lambda^2 P_\Lambda}{R^2} - \frac{\Lambda P_R}{R}\right) = 0.
\end{equation}

The two above conditions will now be treated as gauge-fixing constraints which fix the vector and scalar constraints. The Dirac matrix of the system of the four constraints is given by
\begin{equation}
M_{\alpha\beta}(r,\bar r) = 
	\begin{bmatrix}
	  0 & 0 & -\frac{2m_\text{ADM}}{\chi R^3} + \mathfrak{t}^\text{matt} - \partial_r^2 & 0\\ 
	  0 & 0 & 0 & \partial_r\\ 
	  \frac{2m_\text{ADM}}{\chi R^3}  - \mathfrak{t}^\text{matt} + \partial_{\bar r}^2 & 0 & 0 & -\frac{1}{\chi R^2}\\ 
	  0 & -\partial_{\bar r} & \frac{1}{\chi R^2} & 0\\
	\end{bmatrix}_{\alpha\beta}\delta(r,\bar r)
\end{equation}
on the constraint surface, where $\alpha, \beta$ are indices running through the set of labels $\{H, C, K, \Lambda\}$ denoting correspondingly the constraints \eqref{SSHamiltonianConstraint}$=0$, \eqref{SSVectorConstraint}$=0$, \eqref{SSKrrConstraint} and \eqref{SSLambdaConstraint}, so that in this notation we have in particular $M_{C\Lambda}(r,\bar r)=\{c(r),\ \Lambda(\bar r) - 1\} = \partial_r \delta(r,\bar r)$. The term $\mathfrak{t}^\text{matt}$ denotes the contribution coming from the presence of the cosmological constant and the scalar field. It is given by (we use here the assumptions spelled out under equation \eqref{SSconstraints})
\begin{equation}
	\mathfrak{t}^\text{matt}(r)\delta(r,\bar r) = \frac{1}{\chi R^2(r)}\frac{\delta h^\text{matt}(r)}{\delta \Lambda(\bar r)} - \frac{1}{\chi R(r)}\frac{\delta h^\text{matt}(r)}{\delta R(\bar r)} +  \frac{1}{\chi R^2(r)} h^\textrm{matt}(r) \delta(r,\bar r).
\end{equation}
Since the Dirac matrix is of block form, its formal inversion is of the form
\begin{equation}\label{InverseOfDiracMatrix}
	\begin{bmatrix}
	  0 & A\\
	  -A^\# & B\\
	\end{bmatrix}^{-1}
	=
	\begin{bmatrix}
	  (A^\#)^{-1}BA^{-1} & -(A^\#)^{-1}\\
	   A^{-1} & 0\\
	\end{bmatrix}
	=
	\begin{bmatrix}
	  (A^{-1})^\#BA^{-1} & -(A^{-1})^\#\\
	   A^{-1} & 0\\
	\end{bmatrix},
\end{equation}
where ${}^\#$ denotes a formal transposition of the matrix integral kernel and in the second equality we used the fact that a transposition commutes with an inversion. The block $A$ is diagonal, so its inversion is described in terms of the inverses of the diagonal terms
\begin{equation}
	A^{-1} = 
	\begin{bmatrix}
	  a_2 & 0\\
	  0 & a_1\\
	\end{bmatrix}^{-1} = 
	\begin{bmatrix}
	  a_2^{-1} & 0\\
	  0 & a_1^{-1}\\
	\end{bmatrix}.
\end{equation}
Hence, a formal inversion of the Dirac matrix can be written in the form
\begin{equation}
	(M^{-1})_{\alpha\beta} = 
	\begin{bmatrix}
	0 & -(a_2^{-1})^\#\frac{1}{\chi R^2}a_1^{-1} & -(a_2^{-1})^\# & 0\\
	(a_1^{-1})^\#\frac{1}{\chi R^2}a_2^{-1} & 0 & 0 & -(a_1^{-1})^\#\\
	a_2^{-1} & 0 & 0 & 0\\
	0 & a_1^{-1} & 0 & 0\\
	\end{bmatrix}_{\alpha\beta}.
\end{equation}
The Dirac bracket of two observables ${\cal O}_1$ and ${\cal O}_2$ is formally given by
\begin{equation}
	\{{\cal O}_1,\ {\cal O}_2\}_D = \{{\cal O}_1,\ {\cal O}_2\} - \sum_{\alpha,\beta\in\{H,C,K,\Lambda\}}\{{\cal O}_1,\ \mathfrak{C}_\alpha\}(M^{-1})_{\alpha\beta}\{\mathfrak{C}_\beta,\ {\cal O}_2\},
\end{equation}
where $\mathfrak{C}_\alpha$ denotes the constraint specified by $\alpha$ and the integrations involved in the action of $M^{-1}$ are suppressed here for simplicity. The remaining task is therefore to compute the inverses of $a_1$ and $a_2$.

Finding the inverses of $a_1$ and $a_2$ amounts to solving the following equations
\begin{subequations}\label{SSEqAInDiracBracket}
\begin{align}
	\partial_r N^r &= f,\\
	\partial_r^2 N + (\frac{2m_\text{ADM}}{\chi R^3} - \mathfrak{t}^\text{matt})N &= -g,
\end{align}
\end{subequations}
for $N^r$ and $N$ in terms of $f$ and $g$ respectively. We denoted the unknown functions with the same letters as the shift and lapse because, when looking for lapse and shift functions smearing the constraints which stabilise our gauge fixing constraints, one finds equations of the type above.

The solution of the first equation can be written in the form
\begin{equation}\label{SSSolutionForShift}
	N^r(r) = \int_{r_1}^r d\bar r f(\bar r) + N^r(r_1),
\end{equation}
where the point $r_1$ and the value of $N^r$ at that point can be chosen invoking for example the behaviour of the shift at zero or at infinity (see the discussion of the two possibilities in \cite{BLSII}). For our purposes in the present paper, it is not necessary to specify this function further.

Solving the second equation is a little more complicated. We first find that
\begin{equation}\label{SSSolutionForLapse}
	\begin{bmatrix}
	  N(r)\\
	  N'(r)\\
	\end{bmatrix}
	=
	U(r,r_2)\left(
	\begin{bmatrix}
	  N(r_2)\\
	  N'(r_2)\\
	\end{bmatrix}
	- \int_{r_2}^r d\bar r \, U(r_2,\bar r)
	\begin{bmatrix}
	0\\
	g(\bar r)\\
	\end{bmatrix}
	\right),
\end{equation}
where the resolvent of the homogeneous equation, $U$, is given by
\begin{equation}
	U(r_4,r_3) = \text{Texp}\left(\int_{r_3}^{r_4}d\bar r
	\begin{bmatrix}
	0 & 1\\
	-(\frac{2m_\text{ADM}}{\chi R^3} - \mathfrak{t}^\text{matt}) & 0\\
	\end{bmatrix}
	(\bar r)
	\right),
\end{equation}
where $\text{Texp}$ denotes an ordered exponential. Like in the previous case, the value of the lapse and its derivative at some point $r_2$ have to be specified (the two natural choices again being zero and infinity).

For later reference, let us introduce the following notation. The solutions of equations \eqref{SSEqAInDiracBracket} given by \eqref{SSSolutionForShift} and \eqref{SSSolutionForLapse} will be denoted explicitly referencing the right hand side for which the equation was solved, namely
\be\label{SSNotationSolutions}
	N^r_{[f]},\qquad N_{[g]},
\ee
so that, e.g., $N^r_{[f]}$ is a function, such that
\be
	\partial_r N^r_{[f]} = f.
\ee

\subsubsection{Dirac bracket for massless scalar field}

The key feature we want to stress of the Dirac bracket obtained in the previous section is its non-locality. For the sake of clarity, let us discuss a specific type of scalar field, namely the massless scalar field, and also add a cosmological constant $\Omega$. In such a case, the contribution to the Hamiltonian constraint is given by \cite{RomanoSpherically}
\begin{equation}
	h^\text{matt} = \frac{1}{2\Lambda}\left(\frac{P_\phi^2}{R^2} + R^2\phi'^2\right) + \frac{\chi}{2} \, \Omega \, \Lambda R^2,
\end{equation}
so that the contribution to the Dirac bracket is given by (note, that this is the form on the constraint surface $\mathfrak{C}_\alpha = 0$ for $\alpha\in\{H, C, K, \Lambda\}$)
\begin{equation}
	\mathfrak{t}^\text{matt} = \frac{P_\phi^2}{2\chi R^4} - \frac{3\phi'^2}{2 \chi}.
\end{equation}
We note that the cosmological constant terms cancel in our definition of the spherically symmetric $\mathfrak{t}^\text{matt}$, which was constructed using the vanishing of the Hamiltonian constraint\footnote{\label{ftn:CosConstant}Our definition in the general case, \eqref{eq:tmattgeneral}, retains a contribution coming from the cosmological constant, given by $  - \Omega \,\delta(r,\theta;\bar{r},\bar{\theta})$, since the Hamiltonian constraint was used differently there.}. 

To understand the non-locality of the Dirac bracket, let us consider a Dirac bracket of $\phi$ smeared with a phase-space-independent density $\mu$ (assumed later to be localised at some $r_\mu$, meaning $\mu(r)\sim\delta(r,r_\mu)$) and the same field $\phi$ smeared with a phase-space-independent density $\nu$ (assumed later to be localised at some $r_\nu$). We obtain
\begin{eqnarray}
	\left\{\int\mu\phi,\ \int\nu\phi\right\}_{D}
	&=&-\int drd\bar{r}\left\{\int\mu\phi,\ c(r)\right\}(M^{-1})_{CH}(r,\bar{r})\left\{h(\bar{r}),\ \int\nu\phi\right\}\nonumber\\
	&&-\int drd\bar{r}\left\{\int\mu\phi,\ h(r)\right\}(M^{-1})_{HC}(r,\bar{r})\left\{c(\bar{r}),\ \int\nu\phi\right\}\nonumber\\
	&=&\int drd\bar{r}\mu(r)\nu(\bar{r})\left(\phi'(r)(M^{-1})_{CH}(r,\bar{r})\frac{P_{\phi}}{R^{2}}(\bar{r})+\frac{P_{\phi}}{R^{2}}(r)(M^{-1})_{HC}(r,\bar{r})\phi'(\bar{r})\right)\nonumber\\
	&=&\int drd\bar{r}d\bar{\bar{r}}(\mu\phi')(r)(a_{1}^{-1})^{\#}(r,\bar{r})\frac{1}{\chi R^{2}}(\bar{r})a_{2}^{-1}(\bar{r},\bar{\bar{r}})(\nu\frac{P_{\phi}}{R^{2}})(\bar{\bar{r}})\nonumber\\
	&&-\int drd\bar{r}d\bar{\bar{r}}(\mu\frac{P_{\phi}}{R^{2}})(r)(a_{2}^{-1})^{\#}(r,\bar{r})\frac{1}{\chi R^{2}}(\bar{r})a_{1}^{-1}(\bar{r},\bar{\bar{r}})(\nu\phi')(\bar{\bar{r}})\nonumber\\
	&=&\int drd\bar{r}d\bar{\bar{r}}(\mu\phi')(r)a_{1}^{-1}(\bar{r},r)\frac{1}{\chi R^{2}}(\bar{r})a_{2}^{-1}(\bar{r},\bar{\bar{r}})(\nu\frac{P_{\phi}}{R^{2}})(\bar{\bar{r}})\nonumber\\
	&&-\int drd\bar{r}d\bar{\bar{r}}(\mu\frac{P_{\phi}}{R^{2}})(r)a_{2}^{-1}(\bar{r},r)\frac{1}{\chi R^{2}}(\bar{r})a_{1}^{-1}(\bar{r},\bar{\bar{r}})(\nu\phi')(\bar{\bar{r}}).
\end{eqnarray}
Making use of the discussion about the inverse of the Dirac matrix, and using the notation introduced in \eqref{SSNotationSolutions} we can write
\be\label{SSDiracBracketScalarFields}
	\left\{\int\mu\phi,\ \int\nu\phi\right\}_D =  \int N^r_{[\mu\phi']}\frac{1}{\chi R^2}N_{[\nu\frac{P_\phi}{R^2}]} - \int N_{[\mu\frac{P_\phi}{R^2}]}\frac{1}{\chi R^2}N^r_{[\nu\phi']}.
\ee

\subsubsection{Non-locality of the Dirac bracket in spacetime radial gauge}

Considering $\mu$ and $\nu$ to be localised at $r_\mu$ and $r_\nu$, and analysing the properties of the solutions \eqref{SSSolutionForShift} and \eqref{SSSolutionForLapse}, one can draw a conclusion that in general
\be
	\{\phi(r_\mu),\ \phi(r_\nu)\}_D \neq 0,
\ee
where the non-trivial result of the bracket depends on the canonical data also in points different then $r_\mu$ and $r_\nu$, which constitutes its non-locality. This result contradicts one of the conclusions of \cite{KabatDecodingTheHologram}, however, it is supported by the perturbative analysis of \cite{DonnellyDiffeomorphismInvariantObservables}. We discuss the argument presented in reference \cite{KabatDecodingTheHologram} in comment \ref{com:Kabat} in section \ref{sec:Comments}.

For the sake of clarity of the correspondence with the constructions of observables discussed elsewhere, let us comment that the "anchoring at infinity" construction discussed in \cite{HeemskerkConstructionOfBulk, KabatDecodingTheHologram, DonnellyDiffeomorphismInvariantObservables} corresponds to choosing the points $r_1$ and $r_2$ of the solutions \eqref{SSSolutionForShift} and \eqref{SSSolutionForLapse} to be located at infinity, and specifying the behaviour of the solutions at those points according to the fall-off conditions on the lapse and shift functions. It is important to note that the non-locality presented above cannot be removed by choosing differently the location of $r_1$ and $r_2$, and the values of the solutions of \eqref{SSEqAInDiracBracket} at those points. 

Finally, the computation of the corresponding Dirac bracket in the full theory (without assuming spherical symmetry) yields an analogues result. The non-locality there is present when the two points lie on a common radial geodesic, and is supported along that radial geodesic\footnote{If one would have chosen a Coulomb-type dressing as in \cite{DonnellyDiffeomorphismInvariantObservables}, the conclusion would hold for generic $r_\mu, r_\nu$.}.
We present the computation in the full theory in Appendix \ref{DiracBracketFullTheory}.
In both cases, we see that the non-locality vanishes in the limit of vanishing Newton constant, which is also consistent with the perturbative analysis of \cite{DonnellyDiffeomorphismInvariantObservables}. While the presence of a cosmological constant changes some details, see footnote \ref{ftn:CosConstant}, it does not affect the qualitative result.

\subsection{Non-locality of locally constructed algebras of observables}

A result elucidated by the example in the previous section is that the Dirac bracket for the spacetime radial gauge is generically non-local. Consider a Dirac bracket of two local observables constructed from the canonical fields, where by their locality we mean that the observable is a finite\footnote{For infinite polynomials, one could reconstruct canonical fields in a neighbourhood of the specified point as a Taylor expansion, both in the spatial and the temporal directions.} polynomial of the canonical fields and their spatial derivatives at a given point. From the structure of the Dirac bracket (see Appendix \ref{DiracBracketFullTheory}), we see that the non-local contributions appear if the two observables are supported at points lying on the same radial geodesic. In such a case, regardless of the boundary conditions chosen in the definition of the inverse of the Dirac matrix, there will generically be contributions to the Dirac bracket given by integrals of some combinations of the canonical data along some part of that geodesic (which part of that geodesic depends on the boundary conditions - the choice of the anchoring platform). This fact leads to the conclusion that it is impossible to construct a complete set of local observables having local Dirac brackets using the spacetime radial gauge. The contributions along the involved geodesics cannot be compensated by a specific choice of local observable, which enters the non-local contribution to the Dirac bracket only at the endpoints of the radial geodesic connecting the two points at which the two observables are defined. Moreover, the non-local contributions appear not only in the gravitational sector, but are also present in brackets of matter fields (as shown explicitly), which contradicts a claim of \cite{KabatDecodingTheHologram}, and is in agreement with \cite{DonnellyDiffeomorphismInvariantObservables}.

\section{Comments}\label{sec:Comments}

\begin{enumerate}
	\item From the point of view of Dirac quantisation, i.e. quantisation prior to solving the constraints, it is
 expected that the spacetime radial gauge is problematic: since $\hat K_{rr}$ and $\hat q_{rr}$ cannot commute if the Poisson-algebra is represented properly, the existence of a state $\ket{\psi}$ for which simultaneously $\hat K_{rr} \ket{\psi}= 0$ and $\hat q_{rr} \ket{\psi}= 0$ leads to the standard contradiction observed in the attempt to quantise second class constraints, see e.g. \cite{DiracLecturesOnQuantum}. In other words, existence of a state for which \eqref{eq:Gamma4t}, \eqref{eq:Gamma4r}, and \eqref{eq:Gamma4A} vanish with arbitrary precision, appears to be in conflict with the Heisenberg uncertainty relation.

	\item \label{com:Gauges} One can ask to what extent the spatial radial gauge can be combined with another gauge condition while retaining a local algebra of observables. Of course, this is in general always possible for commuting gauge conditions, such as using an additional matter field as a clock. Another possibility, explicitly involving a geometric clock, would be to follow the ideas of \cite{BSTI} and include a non-minimally coupled scalar field. Due to the mixing of matter and geometric degrees of freedom in the canonical structure of the theory as detailed in \cite{BSTI, BSTII}, geometric gauge conditions such as a constant mean curvature, or fixed value of $K_{rr}$, translate into conformal generators (such as \ref{eq:ConformalGenerator}), also including the scalar field, of the form
	\be
		\alpha(\phi) P^{rr} q_{rr} + \beta(\phi) P^{AB} q_{AB} + \gamma(\phi) \pi_\phi \phi = 0
	\ee 
	for three local functions $\alpha, \beta, \gamma$ of the scalar field, where we also used the gauge fixing $q_{rA}=0$. The ansatz $\tilde q_{rr} = f(\phi) q_{rr}$, $\tilde P^{rr} = f(\phi)^{-1} P^{rr}$, $\tilde q_{AB} = g(\phi) q_{AB}$, $\tilde P^{AB} = g(\phi)^{-1} P^{AB}$ for Dirac observables then gives equations for $f$ and $g$ which can be solved. The spatial radial gauge can now be employed for the rescaled form of the metric, i.e. $\tilde q_{rr} = 1$, whereas $q_{rA}=0$ has already been demanded above. While this gauge is similar to the spacetime radial gauge, in particular if $K_{rr} = 0$ is used as a gauge condition, it is a different gauge condition since from $\tilde q_{rr} = 1$ it does not follow that $q_{rr} = 1$, and thus \eqref{eq:Gamma4t}, \eqref{eq:Gamma4r}, and \eqref{eq:Gamma4A} do not vanish in general.

	\item \label{comment:QFT} Given the problems that we encountered with the spacetime radial gauge, we should ask the question about other suitable gauges for establishing a local quantum field theory limit of quantum gravity. In our point of view, it is most important to choose a gauge in which the algebra of observables is local. Within this class, non-rotating dust as a clock field \cite{BrownDustAsStandard}, or more precisely the proper time elapsing for it, has the advantage that the lapse function is always $1$, so that no non-localities are introduced into the Hamiltonian, as e.g. in the above example, or that it involves taking a square root\footnote{Note that a similar square root resulting from the absolute value of the Hamiltonian as considered in \cite{HusainTimeAndA} in the case of dust is not necessary \cite{SwiezewskiOnTheProperties}.} as e.g. in \cite{RovelliThePhysicalHamiltonian}. Moreover, the back reaction of the dust on the geometry can be made arbitrarily small \cite{BrownDustAsStandard}. Using the dust's proper time also seems to be physically appropriate choice of clock. 
	
	For the spatial diffeomorphisms, the situation is less clear. The spatial radial gauge is very close to the specification of points in usual quantum field theory, whereas it introduces non-localities into the Hamiltonian as the back reaction of matter on the geodesics, scaling with the Newton constant \cite{DuchObservablesForGeneral, BLSII}. However, these non-localities can be controlled at the quantum level as far as a definition of the Hamiltonian is concerned \cite{BLSIII}.
	Additional dust fields usually lead to a square root in the Hamiltonian \cite{BrownDustAsStandard, GieselScalarMaterialReference}.
	
	\item \label{com:Kabat} In \cite{KabatDecodingTheHologram}, an argument was given for why the matter fields should retain their local Poisson brackets. It seems that the error happens around equation (30), where it is stated, in the notation of \cite{KabatDecodingTheHologram}, that $\dot \sigma$ only contains terms proportional to $\dot g_{ij}$. However, there still are time derivatives of the matter fields hiding in the temporal components of the energy momentum tensor, which cannot be translated into the conjugate momenta of the matter fields {\it before} the Legendre transformation. The reasoning for the locality of the matter Poisson brackets therefore appears to be circular at this point.
	
	To address directly a statement made in \cite{KabatDecodingTheHologram}, in which it was said that coupling gravity to matter fields does not modify their brackets, we state that we agree on that point (for the case of minimal coupling). However, as we see from \eqref{InverseOfDiracMatrix}, it is the gauge fixing in which gauge conditions are not mutually Poisson-commuting which leads to modified Poisson (or in fact Dirac) brackets of the matter fields.
	
\end{enumerate}

\section{Conclusion}
\label{sec:Conclusion}

In this paper, we have discussed a spacetime form of the radial gauge, or in other words, the choice of Gau{\ss}ian normal spacetime coordinates in canonical formulations of general relativistic theories. In particular, we showed that the algebra of observables is generically non-local and traced this back to the non-commutativity of the gauge conditions. While this result is certainly interesting for quantum gravity by itself, it has also become of interest recently within the AdS/CFT correspondence. The non-perturbative computation in this paper should settle the debate on the locality of the resulting algebra of observables. Slight deviations of the spacetime radial gauge were suggested which do not suffer from non-locality. It remains however to show that such gauges are also useful within AdS/CFT.

\appendix

\section{Dirac bracket for non-symmetric General Relativity}\label{DiracBracketFullTheory}

In this Appendix, the Dirac bracket for the spacetime radial gauge in the case of non-symmetric General Relativity coupled to a scalar field $\phi$ is presented. A related discussion of the Dirac bracket with gauge conditions \eqref{GaugeConditionMetric} and \eqref{GaugeConditionCurvature} can be found in \cite{DuchMagisterka}. The Dirac bracket of two observables $\mathcal{O}_1$ and $\mathcal{O}_2$ is given by
\begin{multline}
 \{\mathcal{O}_1,\ \mathcal{O}_2\}_D 
 =
 \{\mathcal{O}_1,\ \mathcal{O}_2\} 
 \\
 -
 \sum_{\alpha,\beta=1}^8 \int d r d^2\theta\, d \bar rd^2 \bar{\theta}~
 \{\mathcal{O}_1,\ \mathfrak{C}_\alpha(r,\theta)\} 
 (M^{-1})_{\alpha\beta}(r,\theta;\bar r,\bar{\theta}) 
 \{\mathfrak{C}_\beta(\bar r,\bar{\theta}),\ \mathcal{O}_2\},
\end{multline}
where
\begin{equation}
 \mathfrak{C}_\alpha = (H,C_r,C_A,K_{rr},q_{rr}-1,q_{rA})
\end{equation}
is the set of all constraints in the considered model without assumption of spherical symmetry (we remind the reader that indices $A, B, \ldots$ run over the two angular coordinates)
and
\begin{equation}
 M_{\alpha\beta} = \{ \mathfrak{C}_\alpha,\mathfrak{C}_\beta\}
 =
 \begin{bmatrix}
  0 & A
  \\
  -A^\#& B
 \end{bmatrix}_{\alpha\beta}.
\end{equation}
The entries of the above matrix: $A$ and $B$ are themselves four dimensional matrices (the symbol ${}^\#$ denotes a transposition of a matrix integral kernel):
\begin{equation}\label{DefOfA}
 A(r,\theta;\bar r,\bar\theta)=
 \begin{bmatrix}
  R_{rr}^{(3)} - 2 K_{Ar}K^{Ar}+\mathfrak{t}^\textrm{matt} - \partial_r^2 & 0 & -2 K_{rA}
  \\
  0 & 2\partial_r & \partial_B
  \\
  2 \partial_r K_{rB} & 0 & \partial_r q_{AB}
 \end{bmatrix}\delta(r,\theta;\bar r,\bar{\theta}),
\end{equation}
where the derivatives in the last row act also on the delta, and
\begin{equation}
 B(r,\theta;\bar r,\bar\theta)=
 \begin{bmatrix}
  0 & -\frac{\kappa}{\sqrt{\det q}} & 0 & 0\\
  \frac{\kappa}{\sqrt{\det q}} & 0 & 0 & 0\\
  0 & 0 & 0 & 0 \\
  0 & 0 & 0 & 0 \\ 
 \end{bmatrix}\delta(r,\theta;\bar r,\bar{\theta}).
\end{equation}
In \eqref{DefOfA} we used the notation (assuming that the matter contribution to the Hamiltonian constraint does not depend on the momentum conjugate to the spatial metric)
\begin{equation}
 \mathfrak{t}^\textrm{matt}(r,\theta)\delta(r,\theta;\bar{r},\bar{\theta}) = \frac{\kappa}{\sqrt{q}}\left(\frac{\delta h^{\textrm{matt}}(r,\theta)}{\delta q_{rr}(\bar{r},\bar{\theta})} - q_{AB}\frac{\delta h^{\textrm{matt}}(r,\theta)}{\delta q_{AB}(\bar{r},\bar{\theta})}
 - \frac{1}{2}  h^\textrm{matt}(r,\theta) \delta(r,\theta;\bar{r},\bar{\theta}) \right). \label{eq:tmattgeneral}
\end{equation}
The above Dirac matrix reduces to the one given in the case of spherical symmetry upon an appropriate reduction (note that the Hamiltonian contraint needs to be used to show this).

The inverse of the matrix $M$ is formally given by
\begin{equation}
 M^{-1}
 =
 \begin{bmatrix}
  (A^{-1})^\#BA^{-1} & -(A^{-1})^\#\\
   A^{-1} & 0\\
 \end{bmatrix}.
\end{equation}
In order to find $A^{-1}$ we will solve the equation 
\begin{equation}\label{eq:matrixANM}
 \sum_{\beta=1}^4 \int d\bar r d^2\bar\theta \,A_{\alpha\beta}(r,\theta;\bar r,\bar\theta) N_\beta(\bar r,\bar\theta) = M_\alpha(r,\theta) 
\end{equation}
for $N_\alpha$ ($\alpha=1,\ldots,4$) assuming that the functions $M_\alpha$ ($\alpha=1,\ldots,4$) are given. We use the notation 
\begin{equation}
 N_\alpha = (N,N^r,N^A),~~~~M_\alpha = (M, M^r ,M^A).
\end{equation}
The equation \eqref{eq:matrixANM} is equivalent to the following set of differential equations
\begin{subequations}\label{eq:ANM}
 \begin{align}
  \partial_r^2 N + (-R_{rr}^{(3)} + 2 K_{Ar}K^{Ar}-\mathfrak{t}^\textrm{matt}) N + 2K_{rA} N^A &= -M, \label{eq:ANMa}\\
  2\partial_r N^r + \partial_B N^B &= M_r, \label{eq:ANMb} \\
  \partial_r N_A + 2 \partial_r (K_{r A} N)&= M_A. \label{eq:ANMc}
 \end{align}
\end{subequations}
We first solve equation \eqref{eq:ANMc}
\begin{equation}\label{eq:solutionN_AM}
 N_A(r,\theta) + 2 K_{rA}N(r,\theta) = \int_{r_1}^r d\bar{r}\, M_A(\bar{r},\theta) + N_A(r_1,\theta) + 2 K_{rA}N(r_1,\theta).
\end{equation}
The precise location of the anchoring platform (the value of $r_1$), and the values of the fields in that location are irrelevant for the discussion in the current paper, so we do not specify them further. Using the above result we can write \eqref{eq:ANMa} as follows
\begin{equation}
 \partial_r^2 N - W N  =- \bar{M},
\end{equation}
where $W=R_{rr}^{(3)} + 2 K_{Ar}K^{Ar} + \mathfrak{t}^\textrm{matt}$ and 
\begin{equation}
 \bar{M} = M + 2K^{rA}\int_{r_1}^r d\bar{r}\, M_A(\bar{r},\theta) + 2K^{rA} N_A(r_1,\theta) + 4 K^{rA} K_{rA}N(r_1,\theta).
\end{equation}
Note that $\bar M$ is determined by $M$, $M_A$ and the initial data. We find that
\begin{equation}\label{eq:solutionNM}
 \begin{bmatrix}
  N(r,\theta)\\
  N'(r,\theta)\\
 \end{bmatrix}
 =
 U(r,r_2,\theta)\left(
 \begin{bmatrix}
  N(r_2,\theta)\\
  N'(r_2,\theta)\\
 \end{bmatrix}
 - \int_{r_2}^r d\bar r~ U(r_2,\bar r,\theta)
 \begin{bmatrix}
  0\\
  \bar M(\bar r,\theta)\\
 \end{bmatrix}
 \right),
\end{equation}
where
\begin{equation}
 U(r_2,r_1,\theta) =
   \textrm{Texp}\left( \int_{r_1}^{r_2} d r
  \begin{bmatrix}
    0 & 1 \\
    W(r,\theta) & 0\\
  \end{bmatrix}  \right).
\end{equation}
Using equations \eqref{eq:solutionN_AM} and \eqref{eq:solutionNM}, we determine $N$ and $N^A$ in terms of $M$, $M_A$ and the initial data. In order to find the component $N^r$, we solve equation \eqref{eq:ANMb} and obtain
\begin{equation}\label{eq:solutionN_rM}
 N^r(r,\theta) = \frac{1}{2} \int_{r_3}^r  d\bar{r}\,\left[ M_r(\bar{r},\theta) - \partial_B N^B(\bar{r},\theta)\right] + N^r(r_3,\theta) .
\end{equation}
In what follows, we make the dependence on $(M_\alpha)$ explicit in our notation and write $N_{[M_\alpha]}$, $\vec{N}_{[M_\alpha]}$ for the solution of the equations \eqref{eq:ANM} given explicitly by \eqref{eq:solutionNM},\eqref{eq:solutionN_rM} and \eqref{eq:solutionN_AM} with some arbitrarily fixed initial conditions.

In order to see the non-locality of the Dirac bracket let us calculate the bracket of $\phi$ with itself
\begin{multline}
 \{\phi(r_1,\theta_1),\ \phi(r_2,\theta_2)\}_D =
 \\
 \sum_{\alpha,\beta=1}^4 \int d r d^2\theta d\bar r d^2\bar\theta d\bar{\bar r} d^2\bar{\bar\theta}~ 
 \{\phi(r_1,\theta_1),\ \mathfrak{C}_\alpha(\bar r,\bar \theta)\} 
 ~A^{-1}_{2\alpha}(r,\theta;\bar r,\bar \theta) 
 \\
 \times \frac{\kappa}{\sqrt{\det q(r,\theta)}} 
 ~A^{-1}_{1\beta}(r,\theta;\bar{\bar r},\bar{\bar \theta}) 
 ~\{\mathfrak{C}_\beta(\bar{\bar r},\bar{\bar\theta}),\ \phi(r_2,\theta_2)\}
 \\ 
 -\sum_{\alpha,\beta=1}^4 \int d r d^2\theta d\bar r d^2\bar\theta d\bar{\bar r} d^2\bar{\bar\theta}~ 
 \{\phi(r_1,\theta_1),\ \mathfrak{C}_\alpha(\bar r,\bar \theta)\} 
 ~A^{-1}_{1\alpha}(r,\theta;\bar r,\bar \theta)
 \\
 \times \frac{\kappa}{\sqrt{\det q(r,\theta)}} 
 ~A^{-1}_{2\beta}(r,\theta;\bar{\bar r},\bar{\bar \theta}) 
 ~\{\mathfrak{C}_\beta(\bar{\bar r},\bar{\bar\theta}),\ \phi(r_2,\theta_2)\}
 \\
 =
 \int d r d^2\theta~ 
 N^r_{[\{\phi(r_1,\theta_1),\ \mathfrak{C}_\alpha\}]}(r,\theta) 
 ~\frac{\kappa}{\sqrt{\det q(r,\theta)}} 
 ~N_{[\{\mathfrak{C}_\alpha,\ \phi(r_2,\theta_2)\}]}(r,\theta)
 \\ 
 -\int d r d^2\theta~ 
 N_{[\{\phi(r_1,\theta_1),\ \mathfrak{C}_\alpha\}]}(r,\theta) 
 ~\frac{\kappa}{\sqrt{\det q(r,\theta)}} 
 ~N^r_{[\{\mathfrak{C}_\alpha,\ \phi(r_2,\theta_2)\}]}(r,\theta) .
\end{multline}
Both of the terms generically do not vanish for $(r_1,\theta_1)\neq (r_2,\theta_2)$. In fact, equations \eqref{eq:solutionN_rM}, \eqref{eq:solutionNM}, \eqref{eq:solutionN_AM} show that the kernel $A^{-1}_{\alpha\beta}(r_1,\theta_1;r_2,\theta_2)$ is non-local. To be more precise, it is proportional to $\delta(\theta_1,\theta_2)$, but in general is different than zero for $r_1\neq r_2$ (i.e. it does not vanish for non-coincident points lying on the same radial geodesic).

\section*{Acknowledgements}
This work was partially supported by the Polish National Science Centre grant No.~2011/02/A/ST2/00300 and by Polish National Science Centre grant No.~2013/09/N/ST2/04299.
NB was supported by a Feodor Lynen Research Fellowship of the Alexander von Humboldt-Foundation and during final improvements of this manuscript by the Polish National Science Centre grant No.~2012/05/E/ST2/03308. NB gratefully acknowledges discussions with Antonia Zipfel.


\begin{thebibliography}{100}

\bibitem{DiracLecturesOnQuantum}
P.~A.~M. Dirac, {\em {Lectures on Quantum Mechanics}}.
\newblock Belfer Graduate School of Science, Yeshiva University Press, New
  York, 1964.

\bibitem{HenneauxQuantizationOfGauge}
M.~Henneaux and C.~Teitelboim, {\em {Quantization of Gauge Systems}}.
\newblock Princeton University Press, 1994.

\bibitem{AnishettyGaugeInvarianceIn}
R.~Anishetty and A.~S. Vytheeswaran, ``{Gauge invariance in second-class
  constrained systems},'' {\em Journal of Physics A: Mathematical and General}
  {\bf 26} (1993) 5613--5619.

\bibitem{DittrichPartialAndComplete}
B.~Dittrich, ``{Partial and complete observables for Hamiltonian constrained
  systems},'' {\em General Relativity and Gravitation} {\bf 39} (2007)
  1891--1927, {\tt arXiv:gr-qc/0411013}.

\bibitem{DaporRelationalEvolution}
A.~Dapor, W.~Kami\'nski, J.~Lewandowski and J.~\'Swie\.zewski, ``{Relational Evolution of Observables for Hamiltonian-Constrained Systems},'' {\em Physical Review D} {\bf 88} (2013) 084007, {\tt arXiv:1305.0394 [gr-qc]}.

\bibitem{RovelliWhatIs}
C.~Rovelli, ``{What Is Observable in Classical and Quantum Gravity?},'' {\em Classical and Quantum Gravity} {\bf 8} (1991) 297-316.

\bibitem{DeWittTheQuantizationOf}
{B. S. DeWitt}, ``{The Quantization of geometry},'' in {\em Gravitation: An
  Introduction to Current Research} ({L. Witten}, ed.), (New York),
  pp.~266--381, John Wiley and Sons, 1962.

\bibitem{BrownDustAsStandard}
J.~Brown and K.~Kucha$\check{\text{r}}$, ``{Dust as a standard of space and time in canonical
  quantum gravity},'' {\em Physical Review D} {\bf 51} (1995) 5600--5629, {\tt
  arXiv:gr-qc/9409001}.

\bibitem{GornickaHamiltonianTheoryOf}
J.~Kijowski, A.~Smolski, and A.~G\'ornicka, ``{Hamiltonian theory of
  self-gravitating perfect fluid and a method of effective deparametrization of
  Einstein's theory of gravitation},'' {\em Physical Review D} {\bf 41} (1990)
  1875--1884.

\bibitem{RovelliThePhysicalHamiltonian}
C.~Rovelli and L.~Smolin, ``{The physical Hamiltonian in nonperturbative
  quantum gravity},'' {\em Physical Review Letters} {\bf 72} (1994) 446--449,
  {\tt arXiv:gr-qc/9308002}.

\bibitem{DuchObservablesForGeneral}
P.~Duch, W.~Kami\'nski, J.~Lewandowski and J.~\'Swie\.zewski, ``{Observables for
  general relativity related to geometry},'' {\em Journal of High Energy
  Physics} {\bf 2014} (2014) 77, {\tt arXiv:1403.8062 [gr-qc]}.

\bibitem{HeemskerkConstructionOfBulk}
I.~Heemskerk, ``{Construction of bulk fields with gauge redundancy},'' {\em
  Journal of High Energy Physics} {\bf 2012} (2012) 106, {\tt arXiv:1201.3666
  [hep-th]}.

\bibitem{KabatDecodingTheHologram}
D.~Kabat and G.~Lifschytz, ``{Decoding the hologram: Scalar fields interacting
  with gravity},'' {\em Physical Review D} {\bf 89} (2014) 066010, {\tt
  arXiv:1311.3020 [hep-th]}.

\bibitem{DonnellyDiffeomorphismInvariantObservables}
W.~Donnelly and S.~Giddings, ``{Diffeomorphism-invariant observables and their
  nonlocal algebra},'' {\tt arXiv:1507.07921 [hep-th]}.

\bibitem{BSTI}
N.~Bodendorfer, A.~Stottmeister, and A.~Thurn, ``{Loop quantum gravity without
  the Hamiltonian contraint},'' {\em Classical and Quantum Gravity} {\bf 30}
  (2013) 082001, {\tt arXiv:1203.6525 [gr-qc]}.

\bibitem{DuchAddendumObservablesFor}
P.~Duch, W.~Kami\'nski, J.~Lewandowski and J.~\'Swie\.zewski, ``{Addendum:
  Observables for general relativity related to geometry},'' {\em Journal of
  High Energy Physics} {\bf 2015} 75, {\tt arXiv:1503.07438 [gr-qc]}.

\bibitem{GiddingsHilbertSpaceStructure}
S.~Giddings, ``{Hilbert space structure in quantum gravity: an algebraic
  perspective},'' {\tt arXiv:1503.08207 [hep-th]}.

\bibitem{MaldacenaTheLargeN}
J.~Maldacena, ``{The Large N Limit of Superconformal Field Theories and
  Supergravity},'' {\em Advances in Theoretical and Mathematical Physics} {\bf
  2} (1998) 231--252, {\tt arXiv:hep-th/9711200}.

\bibitem{GubserGaugeTheoryCorrelators}
S.~Gubser, I.~Klebanov, and A.~Polyakov, ``{Gauge theory correlators from
  non-critical string theory},'' {\em Physics Letters B} {\bf 428} (1998)
  105--114, {\tt arXiv:hep-th/9802109}.

\bibitem{WittenAntiDeSitter}
E.~Witten, ``{Anti De Sitter Space And Holography},'' {\em Advances in
  Theoretical and Mathematical Physics} {\bf 2} (1998) 253--291, {\tt
  arXiv:hep-th/9802150}.

\bibitem{AlmheiriBulkLocalityAnd}
A.~Almheiri, X.~Dong, and D.~Harlow, ``{Bulk locality and quantum error
  correction in AdS/CFT},'' {\em Journal of High Energy Physics} {\bf 2015}
  (2015) 163, {\tt arXiv:1411.7041 [hep-th]}.

\bibitem{BLSII}
N.~Bodendorfer, J.~Lewandowski, and J.~\'Swie\.zewski, ``{Loop quantum gravity in
  the radial gauge: Reduced phase space and canonical structure},'' {\em Physical Review D} {\bf 92} (2015) 8, 084041, {\tt
  arXiv:1506.09164 [gr-qc]}.

\bibitem{MintunBulkBoundaryDuality}
E.~Mintun, J.~Polchinski, and V.~Rosenhaus, ``{Bulk-Boundary Duality, Gauge
  Invariance, and Quantum Error Correction},'' {\tt arXiv:1501.06577 [hep-th]}.

\bibitem{BZI}
N.~Bodendorfer and A.~Zipfel, ``{On the relation between reduced quantisation
  and quantum reduction for spherical symmetry in loop quantum gravity},'' {\tt arXiv:1512.00221 [gr-qc]}.

\bibitem{HusainTimeAndA}
V.~Husain and T.~Paw\l owski, ``{Time and a physical Hamiltonian for quantum
  gravity},'' {\em Physical Review Letters} {\bf 108} (2012) 141301, {\tt
  arXiv:1108.1145 [gr-qc]}.

\bibitem{ArnowittTheDynamicsOf}
R.~Arnowitt, S.~Deser, and C.~W. Misner, ``{Republication of: The dynamics of
  general relativity},'' {\em General Relativity and Gravitation} {\bf 40}
  (2008) 1997--2027, {\tt arXiv:gr-qc/0405109}.

\bibitem{KucharGeometrodynamicsOfThe}
K.~Kucha$\check{\text r}$, ``{Geometrodynamics of Schwarzschild black holes},'' {\em Physical Review D} {\bf 50} (1994) 3961--3981, {\tt arXiv:gr-qc/9403003}.

\bibitem{RomanoSpherically}
J. D.~Romano, ``{Spherically Symmetric Scalar Field Collapse: An Example of the Spacetime Problem of Time},'' {\tt arXiv:gr-qc/9501015}.

\bibitem{TorreGravitational}
C. G.~Torre, ``{Gravitational observables and local symmetries},'' {\em Physical Review D} {\bf 48} (1993) R2373(R), {\tt arXiv:gr-qc/9306030}.

\bibitem{BSTII}
N.~Bodendorfer, A.~Stottmeister, and A.~Thurn, ``{On a partially reduced phase
  space quantization of general relativity conformally coupled to a scalar
  field},'' {\em Classical and Quantum Gravity} {\bf 30} (2013) 115017, {\tt
  arXiv:1203.6526 [gr-qc]}.

\bibitem{SwiezewskiOnTheProperties}
J.~\'Swie\.zewski, ``{On the properties of the irrotational dust model},'' {\em
  Classical and Quantum Gravity} {\bf 30} (2013) 237001, {\tt arXiv:1307.4687
  [gr-qc]}.

\bibitem{BLSIII}
N.~Bodendorfer, J.~Lewandowski, and J.~\'Swie\.zewski, ``{Loop quantum gravity in
  the radial gauge II. Quantisation and spherical symmetry},'' {\em (to
  appear)}.

\bibitem{GieselScalarMaterialReference}
K.~Giesel and T.~Thiemann, ``{Scalar Material Reference Systems and Loop
  Quantum Gravity},'' {\tt arXiv:1206.3807 [gr-qc]}.

\bibitem{DuchMagisterka}
P.~Duch, ``{Dirac observables in general relativity constructed by using the Fermi coordinates},'' MSc thesis under the supervision of J.~Lewandowski, University of Warsaw, 2012 {\em (in Polish)}.

\end{thebibliography}

\end{document}